\documentclass[twocolumn,showpacs]{revtex4}
\usepackage{times,xspace}
\usepackage{amsbsy,amssymb,amsmath,bm}
\usepackage{graphicx,color,epsfig,rotate}
\usepackage{fancyheadings}

\def\bbbc{{\mathchoice {\setbox0=\hbox{$\displaystyle\rm C$}\hbox{\hbox
to0pt{\kern0.4\wd0\vrule height0.9\ht0\hss}\box0}}
{\setbox0=\hbox{$\textstyle\rm C$}\hbox{\hbox
to0pt{\kern0.4\wd0\vrule height0.9\ht0\hss}\box0}}
{\setbox0=\hbox{$\scriptstyle\rm C$}\hbox{\hbox
to0pt{\kern0.4\wd0\vrule height0.9\ht0\hss}\box0}}
{\setbox0=\hbox{$\scriptscriptstyle\rm C$}\hbox{\hbox
to0pt{\kern0.4\wd0\vrule height0.9\ht0\hss}\box0}}}}

\newcommand{\ignore}[1]{}
\newcommand{\mComment}[1]{}
\newcommand{\gComment}[1]{}
\newcommand{\jComment}[1]{}
\newcommand{\rComment}[1]{}
\newcommand{\lComment}[1]{}

\renewcommand{\mComment}[1]{\textcolor{blue}{Manny: #1}}
\renewcommand{\gComment}[1]{\textcolor{red}{Gerardo: #1}}
\renewcommand{\jComment}[1]{\textcolor{green}{Jim: #1}}
\renewcommand{\rComment}[1]{\textcolor{magenta}{Ray: #1}}
\renewcommand{\lComment}[1]{\textcolor{purple}{Rolando: #1}}

\begin{document}

\title{Exact Bond Ordered Ground State for the Transition Between the
Band and the Mott Insulator}
\author{C. D. Batista$^1$ and A. A. Aligia$^2$}
\address{$^1$Center for Nonlinear Studies and Theoretical Division,
Los Alamos National Laboratory, Los Alamos, NM 87545\\
$^2$ Centro At\'omico Bariloche and Instituto Balseiro, Comisi\'on Nacional 
de Energ\'{\i}a
At\'omica, 8400 Bariloche, Argentina}

\date{Received \today }

\begin{abstract}
We derive an effective Hamiltonian $H_{eff}$ for an ionic
Hubbard chain, valid for $t\ll U,\Delta $, where $t$ is the
hopping, $U$ the Coulomb repulsion, and $\Delta$ the charge transfer
energy. $H_{eff}$ is the minimal model for describing the transition from
the band insulator (BI) ($\Delta -U\gg t$) and the Mott insulator (MI) 
($U-\Delta \gg t$). Using spin-particle transformations (Phys.
Rev. Lett. \textbf{86}, 1082 (2001)), we map
$H_{eff}(U=\Delta )$ into an SU(3) antiferromagnetic Heisenberg model whose
exact ground state is known. In this way, we show rigorously that a
spontaneously dimerized insulating ferroelectric phase appears in the
transition region between the BI and MI.
\end{abstract}


\maketitle

It is well known that the insulating state can have different origins.
The simplest case is the BI since it is one of the
possible solutions for non-interacting electrons moving in a periodic potential.
In particular, the BI occurs when the number of particles
per unit cell is even, so the bands are either full or empty.
The other traditional example of insulating state is the MI. In this case, the charge localization is just a consequence of the local Coulomb interaction $U$ between the 
electrons. The charge gap of the BI is just the band gap. In the case of the
MI, this gap is a function of the Coulomb repulsion and goes asymptotically  to
$U$ in the strong coupling limit. These two insulating states have completely different 
properties. For instance, the BI's are paramagnets while the MI's,
in general, are antiferromagnets. This observation rises the following question: what happens 
when a system evolves continuously from the BI to the MI state?

To find an answer to this question has been the main motivation for 
studying the Ionic Hubbard model (IHM) during the last ten years.
The IHM was originally proposed in the 80's 
to describe the neutral-ionic transition in mixed stack charge-transfer
organic crystals \cite{Hubbard,Nagaosa}. This model is a Hubbard Hamiltonian on a bipartite 
lattice with different diagonal energies for the two sublattices. The difference 
between both energies is $\Delta$. During the 90's, there was a renewed interest in 
the IHM due to the potential applications to the description of the 
ferroelectric (FE) perovskites \cite{Egami,Resta,Fabrizio}.
At half filling, the ground state of the IHM is an ionic or BI for $\Delta \gg U$
and a MI for $U \gg \Delta$. Field theory arguments in one dimension
pointed out the existence of an intermediate  bond ordered insulating  
(BOI) phase for $\Delta \ll U \ll t$ 
\cite{Fabrizio}. For $t\ll U, \Delta$,
perturbation theory clearly describes the BI ($t \ll \Delta-U$)
and the MI ($t \ll U-\Delta$) \cite{Nagaosa}, including the 
charge dynamics of the latter \cite{aligia}. However, perturbation 
theory diverges in the transition region and no insight is provided 
for the BOI. The numerical solutions of finite chains
\cite{Torio,Zhang,Manmana,Kampf} have difficulties and contradictory conclusions 
were reported. The reason 
will become clear after deriving the main result of the present paper.

Contrary to the cases of the BI and the MI
states, no exact solution having long range bond ordering 
is known for the transition regime. Finding an
exact ground  state is not only crucial to prove 
the existence of the BOI phase rigorously, but also to understand its
microscopic origin and fundamental properties. This result becomes even more important
if we consider that the BOI phase is an electronically induced
spin-Peierls instability that generates a FE state out of the 
spin-singlet dimer pairs. In particular, a bond ordered FE state was observed 
in the pressure-temperature phase diagram of the prototype compound,
tetrathiafulvalene-$p$-chloranil \cite{Lemee,Horiuchi}. In addition,
as it was pointed out by Egami {\it et al.} \cite{Egami}, 
the microscopic origin
of the FE transition in covalent perovskite oxides
like BaTiO$_3$ is still unclear. It is known that a picture 
based on static Coulomb interactions and the simple shell model is inadequate to describe
some FE properties \cite{Hippel}. The exact result presented here demonstrates
that when an ionic insulator gets close to  a charge transfer instability (strong covalency),
an electronic mechanism for ferroelectricity takes place.

In this paper, we first derive an effective Hamiltonian, $H_{eff}$, for the 
limit $U\gg t$ and $\Delta \gg t$ of an extended IHM that includes a
nearest neighbor Coulomb repulsion $V$. By means of the 
generalized spin-particle transformations introduced 
in Refs~\cite{bat1,bat2,bat3}, we map $H_{eff}$ into an
anisotropic SU(3) antiferromagnetic Heisenberg model that 
becomes isotropic for $U=\Delta$ and particular values of the 
other parameters. The isotropic model is exactly solvable \cite{Parkinson,Barber,Klumper} 
and is also equivalent to the biquadratic $S=1$ Heisenberg model. 
The exact ground state is a dimerized spin system that becomes
a BOI when translated back to the original fermionic variables.

We will start considering an IHM with an additional 
Coulomb interaction $V$ between nearest neighbors:
\begin{eqnarray}
H &=& -t \sum_{i,\sigma} (f^{\dagger}_{i+1 \sigma} f^{\;}_{i \sigma}+ {\rm H.c.})
+ \frac{\Delta}{2} \sum_{i} (-1)^{i} n_{i} 
\nonumber \\
&+&  U \sum_{i} (n_{i \uparrow}-\frac{1}{2}) (n_{i\downarrow}-\frac{1}{2})
\nonumber \\
&+& V \sum_{i} (n_{i}-1)(n_{i+1}-1),
\label{Hamil}
\end{eqnarray}
where $n_{i \sigma}= f^{\dagger}_{i \sigma}f^{\;}_{i \sigma}$,
$n_{i}=\sum_{\sigma} n_{i \sigma}$, and $t>0$.
Note that $\Delta>0$ is the diagonal energy difference between sites 
in different sublattices.

\begin{figure}[htb]
\vspace*{-2.5cm}
\includegraphics[angle=0,height=9cm,width=8.5cm,scale=1.0]{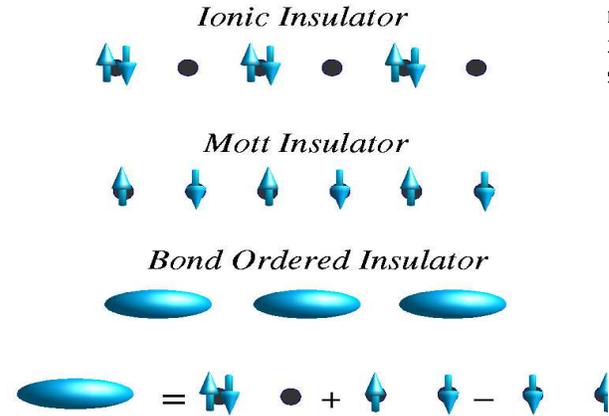}
\vspace*{-0.0cm}
\caption{
\label{fig1}
Schematic plot of the different ground states of $H_{eff}$.}
\end{figure}

We will only consider the half-filled case $\rho=1$, i.e.,
one particle per site. If $t=0$, the ground state of $H$ 
is a non-degenerate BI for $U<\Delta+2V$. There 
is only \emph{one} low-energy state per site: the
odd sites are doubly occupied and the even sites are empty (see Fig.\ref{fig1}).
For $U>\Delta +2V$ the ground state is a degenerate MI (one particle per site) 
and the low-energy subspace has \emph{two} states per site due to 
the spin degeneracy. When $t$ is finite and small, far from the transition region 
the low-energy Hamiltonian of the MI is a Heisenberg model  and the degeneracy is
lifted in favor of a spin-density wave \cite{Nagaosa,aligia} (see Fig.\ref{fig1}). 
However, to describe the transition region, we need
to include \emph{three} states per site in the low-energy subspace 
$\mathcal{H}_{0}$. In order to construct an effective Hamiltonian in $\mathcal{H}_{0}$, 
it is convenient to perform an electron-hole transformation for the odd sites:
\begin{eqnarray}
c_{i\uparrow }^{\dagger } &=&-f_{i\downarrow }^{\;},\;\;\;c_{i\downarrow
}^{\dagger }=f_{i\uparrow }^{\;},\;\;\text{for odd }i  \nonumber \\
c_{i\sigma }^{\dagger } &=&f_{i\sigma }^{\dagger },
\;\;\;\;\;\;\;\;\;\;\;\;\;\;\;\;\;\;\;\;\;\;\;
\text{for even }i  
\label{eh}
\end{eqnarray}
After this transformation, $H$ becomes invariant under a translation 
of one lattice space ($i \rightarrow i+1$).
The low energy subspace ${\cal H}_0$ is now defined as the set of states
with no double occupancy on any site. This constraint can
be incorporated  by defining
the constrained fermion operators:
${\bar c}^{\dagger}_{i\sigma}=c^{\dagger}_{i \sigma} 
(1-c^{\dagger}_{i {\bar \sigma}}c^{\;}_{i {\bar \sigma}})$, and 
${\bar c}^{\;}_{i\sigma}= ({\bar c}^{\dagger}_{i\sigma})^{\dagger}$.
In addition, to connect our low energy theory with spin
Hamiltonians (see below), we introduce the following transformation:
\begin{eqnarray}
\nonumber
&{\bar c}^{\dagger}_{i \downarrow}& \rightarrow -{\bar c}^{\dagger}_{i \downarrow}
\;\;\;\;\text{for} \;\;i=4n\;\;\text{and}\;\;i=4n+1
\\
&{\bar c}^{\dagger}_{i \uparrow}& \rightarrow -{\bar c}^{\dagger}_{i \uparrow}
\;\;\;\;\text{for}\;\; i=4n+2 \;\;\text{and}\;\;i=4n+1
\label{gauge}
\end{eqnarray}
The high-energy subspace $\mathcal{H}_{1}$ consists of all states
which have at least one double occupied site. The 
only term of $H$ that mixes $\mathcal{H}_{0}$ and $\mathcal{H}_{1}$ is 
the hopping term $t$. By means of a second order canonical transformation that eliminates 
the terms proportional to $t$, we obtain the following effective Hamiltonian 
for the low energy spectrum of $H$:
\begin{eqnarray}
H_{eff} &=& t \sum_{i,\sigma} ({\bar c}^{\dagger}_{i,\sigma} {\bar 
c}^{\dagger}_{i+1,\bar \sigma}+H.c.)+ \frac{\Delta-U}{2} \sum_{i} {\bar n}_i
\nonumber \\
&+& J \sum_{i} (s^{z}_i s^{z}_{i+1}- s^{x}_i  
s^{x}_{i+1}
- s^{y}_i s^{y}_{i+1}-\frac{1}{4} {\bar n}_i{\bar n}_{i+1})
\nonumber \\
&-& V \sum_{i} (1-{\bar n}_{i})(1-{\bar n}_{i+1}),
\label{hpp}
\end{eqnarray}
where ${\bar n}_i=\sum_{\sigma} {\bar c}^{\dagger}_{i \sigma}{\bar c}^{\;}_{i \sigma}$
and $s^{\alpha}_i= 1/2 \sum_{\tau,\tau'} {\bar c}^{\dagger}_{i \tau} 
\sigma^{\alpha}_{\tau\tau'}{\bar c}^{\;}_{i \tau'}$ with $\alpha=\{x,y,z\}$
($\sigma^{\nu}$ are the Pauli matrices).
The exchange interaction $J=2t^2/(U+\Delta-V)$ comes from a second order 
process in the hopping $t$. In this derivation we have neglected the 
second order three-site hopping term. The negative sign for the $xx$ and $yy$ interactions
is just a consequence of the gauge transformation of Eq.~(\ref{gauge}). 

If $t \sim V$ and $\Delta-U \gg t$, the ground 
state of $H_{eff}$ is the empty state that corresponds to the BI in the original language. 
For $U-\Delta\gg t$, the ground state has one particle per site.
By eliminating states with empty sites with another canonical transformation, 
$H_{eff}$ reduces to the Heisenberg model that describes the strong coupling limit
of the MI. We are interested in the transition 
regime $U\sim \Delta $. To analyze this case, we will allow $J$ to vary
independently of the other parameters. In other words, we will consider 
$H_{eff}$ as an independent minimal model for describing the transition
between the BI and the MI.


To exploit the symmetries of $H_{eff}$, it is convenient to use 
the generalized Jordan-Wigner transformations introduced in Refs.~\cite{bat1,bat2,bat3}.
In particular, for this case it is appropriate rewrite $H_{eff}$ 
in terms of the hierarchical SU(3) language \cite{bat3}. To this end,
we need to map the constrained fermion operators 
in one sublattice (say $A$) into SU(3) spins
in the fundamental or ``quark'' representation \cite{bat2,bat3}:
\begin{eqnarray}
{\cal S}({j})= \begin{pmatrix} \frac{2}{3} - \bar{n}_{j}
& K^{\dagger}_j\bar{c}^{\;}_{j \uparrow} & K^{\dagger}_j \bar{c}^{\;}_{j 
\downarrow} \\
\bar{c}^{\dagger}_{j \uparrow}K^{\;}_j&\bar{n}_{{j} \uparrow} 
-\frac{1}{3}&
\bar{c}^{\dagger}_{ j \uparrow} \bar{c}^{\;}_{ j \downarrow} \\
\bar{c}^{\dagger}_{j \downarrow} K^{\;}_j &\bar{c}^{\dagger}_{ j \downarrow}
\bar{c}^{\;}_{ j \uparrow}&\bar{n}_{ j \downarrow}-\frac{1}{3}
\end{pmatrix} \ ,
\label{spinsu3p}
\end{eqnarray}
where $K^{\;}_j$ is the kink operator \cite{bat1}:
\begin{equation}
K^{\;}_{j} = \exp[i \pi \sum_{\substack{k<j}}
\ \bar{n}_k] \ .
\end{equation}
that transmutes the statistics. The components ${\cal S}^{\mu \nu}$ are 
generators of the $su(3)$ algebra with commutation relations  $[{\cal S}^{\mu \mu'}({j}),{\cal S}^{\nu
\nu'}({j})]= \delta_{\mu' \nu} {\cal S}^{\mu \nu'}({j})-\delta_{\mu \nu'} {\cal S}^{\nu \mu'}({ j})$.
We also make use of the conjugated or ``anti-quark'' representation:  
\begin{eqnarray}
\tilde{\cal S}(j)= \begin{pmatrix} \frac{2}{3} - \bar{n}_{j}
&-\bar{c}^{\dagger}_{j \downarrow} K^{\;}_j & -\bar{c}^{\dagger}_{ j 
\uparrow} K^{\;}_j
\\ - K^{\dagger}_j \bar{c}^{\;}_{ j \downarrow}&\bar{n}_{ j \downarrow}
-\frac{1}{3}& \bar{c}^\dagger_{{j} \uparrow} \bar{c}^{\;}_{ j
\downarrow} \\ - K^{\dagger}_j \bar{c}^{\;}_{{j} 
\uparrow}&\bar{c}^\dagger_{
j \downarrow} \bar{c}^{\;}_{j \uparrow}&\bar{n}_{ j \uparrow}-\frac{1}{3}
\end{pmatrix} \,
\label{spinsu}
\end{eqnarray}
to describe the degrees of freedom of the $B$ sublattice.
Eqs.~(\ref{spinsu3p}) and (\ref{spinsu}) are generalizations of the 
Jordan-Wigner transformations
to SU(3) spins \cite{bat2,bat3}. The resulting 
$H_{eff}$ is as an anisotropic SU(3) Heisenberg model with an 
applied ``magnetic field'':
\begin{eqnarray}
H_{eff} = \sum_{i \in A ,\mu,\nu} J_{\mu \nu}
{\cal S}^{\mu \nu}(i) {\tilde {\cal S}}^{\nu \mu}(i+1) - 
B \sum_{i} { \cal S}^{00}(i),
\label{su3}
\end{eqnarray}
with $J_{00}=-V$, $J_{01}=J_{02}=-t$,  $J_{11}=J_{12}=J_{22}=-J/2$, 
$J_{\mu \nu}=J_{\nu \mu}$, 
and 
$B=2V/3-J/3+U/2-\Delta/2$. Note that this model 
connects SU(3) spins in the $A$ sublattice with the conjugate SU(3)
spins (anti-quark representation) in the $B$ sublattice.
Let us now consider $V=J/2=t$ and $U=\Delta$. 
For this line in the space of parameters $H_{eff}$ is an {\it isotropic SU(3) antiferromagnetic Heisenberg model}
that is invariant under staggered conjugate SU(3) rotations, ${\cal R}$ and ${\cal R}^{\dagger}$,
on sublattices $A$ and $B$ respectively.
This  model  is integrable \cite{Parkinson,Barber,Klumper} 
and the exact solution is a spin-dimerized ground state. In the original language this 
is equivalent to saying that the charge and the spin are both dimerized 
(bond ordering). The exact ground state energy per site is $e_0/t=-1.796864...$.
The value of the gap is ${\tilde \Delta/t=0.173178}$, a rather small value, and the correlation 
length $\xi=21.0728505...$ is very large \cite{Klumper}. This explains the 
numerical difficulties for identifying this phase.

To make clear the relation between the dimerization in SU(3) and 
the bond ordering of the fermionic variables, we just need to translate
the corresponding order parameter from the SU(3) language back to our original fermionic language.
The spin-dimer SU(3) order parameter is \cite{Xian,Solyom}:
\begin{equation}
D=|h_{i-1,i}-h_{i,i+1}|
\label{op}
\end{equation}
where $h_{i-1,i}=|i-1,i\rangle \langle i,i-1|$ is a projector on the SU(3) singlet spin state,
$|i-1,i\rangle$, at the bond $(i-1,i)$.  Note, in addition, that $h_{i-1,i}$ is  the isotropic 
SU(3) Heisenberg Hamiltonian for the same bond. In the fermionic language,  $|i-1,i\rangle$ 
has the following expression: 
\begin{equation}
|i-1,i\rangle = \frac{1}{\sqrt{3}}  (1- {\bar c}^{\dagger}_{i-1 \uparrow}{\bar c}^{\dagger}_{i \downarrow}
-{\bar c}^{\dagger}_{i-1 \downarrow} {\bar c}^{\dagger}_{i \uparrow}) |0 \rangle
\label{singlet}
\end{equation}
In terms of the original $f$-fermions [see Eqs.(\ref{eh}) and (\ref{gauge})], this is a linear 
combination of an on-site singlet and a nearest-neighbor singlet state that is illustrated 
at the bottom of Fig.~\ref{fig1}. Xian showed that $D=0.4216D_{0}$ where $
D_{0}$ is the value of $D$ for a perfect dimerized state \cite{Xian}. Replacing 
(\ref{singlet}) in (\ref{op}), it becomes clear
that $D$ is a BOI order parameter. 
From a calculation of the charge Berry phase \cite{Torio} we obtain that the shift
in polarization of the two possible perfect dimerized states (see the bottom of Fig.\ref{fig1})
relative to the MI is $\pm e/6$.
Eq.~(\ref{singlet}) shows that the dimer formation
is just a consequence of the charge transfer instability between the two sublattices.
In this sense, this exact ground state unveils the fundamental role of {\it covalency} for 
the stabilization of a bond ordered FE state \cite{Lemee,Horiuchi,Egami}.
This is not only relevant to describe one dimensional systems like 
tetrathiafulvalene-$p$-chloranil \cite{Lemee,Horiuchi}, but also helpful to gain a 
deeper understanding of the FE transition of covalent perovskites \cite{Egami}.

Using another set of transformations  that connect the constrained fermions
with SU(2) $S=1$ spins \cite{bat1}:
\begin{eqnarray}
S^+_j &=& \sqrt{2} \ (\bar{c}^{\dagger}_{j \uparrow} \ K_j + K_j^{\dagger} \
\bar{c}^{\;}_{j \downarrow}) , 
\nonumber \\ 
S^-_j &=& \sqrt{2} \ (K_j^{\dagger} \ \bar{c}^{\;}_{j \uparrow} +
\bar{c}^{\dagger}_{j \downarrow} \ K_j) , 
\nonumber \\ 
S^z_j \ &=& \bar{n}_{j \uparrow} - \bar{n}_{j \downarrow}, 
\end{eqnarray}
we can also write $H_{eff}(V=J/2=t)$ as an $S=1$ biquadratic 
Heisenberg model \cite{bat2} with a single-ion anisotropy:
\begin{equation} 
H_{eff}(V=J/2=t) = -t \sum_{i} 
\left( {\bf S}_{i} \cdot {\bf S}_{i+1} \right)^2 
+ E \sum_{i} (S^z_i)^2,
\label{spin1}
\end{equation}
where $E=(\Delta - U)/2$. The strength of the  anisotropy term
is determined by the difference $\Delta-U$. If $E$ is large and 
positive ($\Delta \gg U$), the spin system has an easy plane anisotropy.
Each site is most of the time in the $S^z=0$ state, which means that 
the magnetization is perpendicular to the $z$-axis. The ground state
is non-degenerate (there is no broken symmetry) and corresponds, in 
the original language, to the {\it band insulator}. If $E$ is large and 
negative ($U \gg \Delta$), the system has a strong  easy axis anisotropy
and each site is in the $S^z=\pm 1$ state, i.e., the local magnetization is
parallel to the $z$-axis. The ground state is critical due to the antiferromagnetic 
correlations which characterize the MI. In between, for $U=\Delta$,
we have demonstrated that there is a dimerized state which corresponds 
to the BOI. In terms of the original variables, the strong 
quantum fluctuations that appear in the proximity of the  charge transfer instability break the
$Z_2$ inversion symmetry by increasing the strength of one bond relative to 
the next one. 

The $S=1$ version of $H_{eff}$ also provides a simple way of studying the low
energy excitations of our exact bond ordered state. The excitations of
a dimerized spin $1/2$ chain are spinons that carry a spin $S=1/2$.
In the same way,  the excitations of our $S=1$ dimer state are $S=1$ spinons.
Each spinon is a soliton or anti-phase boundary for the $Z_2$ spin-dimer 
order parameter. The two regions with opposite dimerization are separated 
by a local $S=1$ defect which is attached to the anti-phase boundary.
In terms of the original language, the $S_z=0$ spinons correspond to 
$s=0$ solitons (charge excitations), while the $S_z=\pm 1$ spinons are
$s=1/2$ solitons (spin and charge excitations). 
From Eqs. (\ref{eh}) and (\ref{spin1}), the total charge  
operator relative to half-filling is  
$Q=\sum_{i\in A} (S^{z}_i)^2 -\sum_{i\in B} (S^{z}_i)^2$.
For a general dimerized solution with arbitrary $E$ we have on each site:  
$\langle (S^{z}_i)^2 \rangle=1- \alpha$ with $\alpha=1/3$ for the exact 
solution at $E=0$. When the defect is localized on site $j$ 
(the extension of the defect does not affect the charge or spin of 
the excitations because they are topological invariants) we have
$\langle (S^{z}_j)^2 \rangle=1,0$. Since the defect spearates two 
regions with opposite dimerization, it is easy to check that
the $s=0$ solitons (anti-solitons) have charge $Q= \pm (1-\alpha)$ while the
charge of the $s=1/2$ solitons(anti-solitons) is $Q= \pm \alpha$.
For $E=0$, both excitations are degenerate as 
a consequence of the SU(3) invariance.
These excitations coincide with those obtained by Fabrizio {\it et al} \cite{Fabrizio}, 
who treat the bosonized IHM as a phenomenological Ginzburg-Landau energy functional. 
The magnitud $\alpha$ is proportional to their  jump in the charge field.
In particular, the $s=1/2$ excitations interpolate between an ordinary
electron near the BI-BOI boundary ($\alpha=1$, $E >0$) and a spinon near the 
the BOI-MI boundary ($\alpha=0$, $E < 0$).

In order to obtain an exact ground state of $H_{eff}$ we used a value of $J=2t$
which is beyond the region allowed by perturbation theory. Therefore, to connect our 
exact solution with the IHM, it is important to discuss the effect of reducing the value of $J$.
In addition, since most of the previous papers do not include the $V$ term, it is also
important to analyze its effect. A simple first order estimation of the 
energy change for the three competing phases, $\delta E_{BI}=\delta U/2+t$,  
$\delta E_{MI}=2t ln2$, $\delta E_{BOI}=\delta U/6-5e_0t/9$, when $J$ and $V$ 
are reduced to zero, indicates that the BOI survives  if the difference $U-\Delta$
is simultaneously increased to a value of order $t$. This conclusion is 
supported by different numerical results \cite{Torio,Zhang,Manmana} that 
report a bond ordered ground state of $H(V=0)$ for the same region of parameters 
($U-\Delta \sim t$ and $U,\Delta \gg t$). These observations suggest that 
our exact solution is continuously connected with the BOI phase which was
numerically found in the IHM.


In summary, we have derived an effective low energy Hamiltonian, $H_{eff}$, for 
the $U,\Delta \gg t$ limit of the IHM. $H_{eff}$ is a minimal model to describe 
the BI to MI transition. Its simple form and the fact that it operates in a  reduced
Hilbert space of local dimension $D=3$ provide a new framework to understand 
this transition. Using the spin-particle transformations introduced in Refs.~\cite{bat2,bat3},
we mapped $H_{eff}$ into an anisotropic SU(3) antiferromagnetic Heisenberg
model. By increasing the value of $J$ beyond the region allowed by perturbation theory, 
we have shown that there is an exactly solvable SU(3) invariant point for $U=\Delta$ and $V=J/2=t$.
In this way, we demonstrated the existence of a bond ordered phase 
for the transition regime between the BI and the MI. 
The large value of the correlation length $\xi=21.0728505...$ explains
the numerical difficulties for detecting this phase 
in finite size systems. This exact solution provides an answer 
to the question that was formulated in the introduction: in the transition regime
the system resolves the competition between the BI and the MI by 
creating a rather local resonance which can be
visualized on each dimer as a linear combination of a unit cell of the BI 
and a nearest-neighbor singlet that is related with the MI 
(see Fig~\ref{fig1}). One of the most important physical consequences of this non-trivial
phenomenon is the generation of a new mechanism for ferroelectricity 
\cite{Egami} that should be relevant for ionic insulators which are close to a charge-transfer
instability. This result should motivate a careful reexamination of  
the ferroelectricity in covalent materials.

In addition to the SU(3) mapping, we used a second spin-particle transformation 
\cite{bat1} that maps the constrained fermions into $S=1$ SU(2) spins. In this 
new language $H_{eff}(V=J/2=t)$ is a biquadratic Heisenberg model with a single-ion anisotropy 
along the $z$-axis which is proportional to $\Delta-U$. The transitions between 
the BI, BOI, and MI phases have been re-interpreted in the $S=1$ language. 
We have also shown that the $S=1$ version of $H_{eff}$ provides the most 
natural frame to understand the low energy excitations of the BOI. These excitations
are solitons that carry a fractional charge and spin $s=0$ or $s=1/2$. The relevance of these
results illustrates the elegance and the potential of the generalized spin-particle transformations introduced in Refs.~\cite{bat1,bat2,bat3}.

We thank J. E. Gubernatis for a careful reading of the manuscript.  
This work was sponsored by the US DOE under contract
W-7405-ENG-36, PICT 03-06343 of ANPCyT. A. A. Aligia
is partially supported by CONICET.

\end{document}